\documentclass[12pt]{article}

\usepackage{latexsym}
\usepackage{amssymb}
\usepackage{amsfonts}

\usepackage{color,graphicx} 
\definecolor{brownn}{cmyk}{0,1,1,0.5}
\definecolor{bluen}{rgb}{.1 ,0, .8}

\usepackage[hypertexnames=false,
colorlinks,linkcolor=bluen,citecolor=brownn,urlcolor=brownn]{hyperref}
\graphicspath{{images/}}

\def\nc#1{\newcommand{#1}}
\def\rnc#1{\renewcommand{#1}}

\def\a{\alpha}

\nc{\g}{\gamma}
\rnc{\d}{\delta}
\nc{\D}{\Delta} 
\nc{\e}{\eta}
\nc{\ep}{\epsilon}

\nc{\ve}{\varepsilon}
\nc{\G}{\Gamma}

\nc{\la}{\lambda}
\nc{\La}{\Lambda}
\nc{\om}{\omega}
\nc{\Om}{\Omega}
\nc{\vphi}{\varphi}
\nc{\si}{\sigma}
\nc{\Si}{\Sigma}
\rnc\th{\theta}
\nc\Th{\Theta}
\nc{\z}{\zeta}

\def\cO{{\cal O}}

\nc{\got}[1]{\mathfrak{#1}} 
\def\dim{{\rm dim}}

\def\det{{\rm det}}

\nc\im{{\rm Im}\, }
\nc\re{{\rm Re}\, }
\def\tr{{\rm tr}}

\nc{\Rt}{{\tilde R}}
\nc{\CC}{{\mathbb C}}
\nc\one{{\mathbb I}} 
\nc{\RR}{{\mathbb R}}
\nc{\HH}{{\mathbb H}}
\nc{\NN}{{\mathbb N}}
\nc{\ZZ}{{\mathbb Z}}
\nc{\MM}{{\mathbb M}}

\nc{\eql}{\eqalign}
\nc{\dis}{\displaylines}
\nc{\ce}{\centerline}
\nc{\hf}{\hspace{\fill}}
\nc{\hs}{\hspace*}
\nc{\vs}{\vskip .3cm}
\nc{\non}{\nonumber\\}
\def\nn{\nonumber}

\nc{\noi}{\noindent}

\nc\ov[1]{\overline{\mbox{\raisebox{0pt}[1.8ex][0pt]{$#1$}}}}

\nc{\p}{\partial}
\nc{\na}{\nabla}

\def\ox{\otimes}

\nc\vev[1]{\ensuremath{\langle #1\rangle} {}}
\nc\vac{\ensuremath{|0\rangle} }

\nc{\lan}{\langle}
\nc{\ran}{\rangle}

\nc{\beq}[1][]{\begin{equation}\label{#1}}
\nc{\eeq}{\end{equation}}
\nc{\beqa}{\begin{eqnarray}}
\nc{\eeqa}{\end{eqnarray}}
\nc{\beqas}{\begin{eqnarray*}}
\nc{\eeqas}{\end{eqnarray*}}
\nc{\barr}{\begin{array}}
\nc{\earr}{\end{array}}
\nc{\bmin}[2][]{\begin{minipage}[#1]{#2}}
\nc\emin{\end{minipage}}
\nc{\ben}{\begin{enumerate}}
\nc{\een}{\end{enumerate}}
\nc{\bit}{\begin{itemize}}
\nc{\eit}{\end{itemize}}
\rnc{\-}{\item}

\nc{\sfrac}[2]{{\mbox{\large $\frac{#1}{#2}\,$}}}
\nc\half{\sfrac{1}{2}}
\nc\nfrac[2]{\mbox{\small $\frac{#1}{#2}$}}

\nc{\eq}[1]{\stackrel{#1}{=}}

\rnc\to{\ensuremath{\rightarrow\;}}

\nc{\oto}{\leftrightarrow}
\nc{\ot}{\leftarrow}
\nc\To{\Rightarrow}

\def\refeq#1{{(\ref{#1})}}

\nc{\twovec}[2]{\left( \!\!
\begin{array}{c} #1\\  #2 \end{array}\!\!\right)}
\nc{\stwovec}[2]{\bigg( \!\!
\begin{array}{c} \mbox{\raisebox{.7ex}{$#1$}}\\[-.3cm]  #2
\end{array}\!\!\bigg)}
\nc{\twomat}[4]{\left(\!\! \begin{array}{cc} #1&#2\\ 
#3&#4\end{array}\!\! \right)}
\nc{\stwomat}[4]{\bigg(\!\!\barr{cc}
#1&\!\!\!\!#2\\[-.2cm]#3&\!\!\!\!#4
\earr\!\!\bigg)}

\rnc{\sec}[1]{\section{\textcolor{bluen}{\large #1}}}
\nc{\ssec}[1]{\subsection{#1}}
\nc{\sssec}[1]{\subsubsection{\sc #1}}

\makeatletter
\@addtoreset{equation}{section}

\makeatother

\nc\rQ[1][]{\ensuremath{{\textcolor{red}{{\leftarrow(?)}}\mbox{{\footnotesize #1}}} } }
\nc\lQ{\ensuremath{{\textcolor{red}(?)\rightarrow}} }

\usepackage{amsmath}

\def\LL{\mathbb{L}}

\def\one{\mathbb{I}}
\nc{\A}[2]{A^{(#2)}_{#1}}
\rnc{\Im}{\mbox{Im}}\rnc{\Re}{\mbox{Re}}
\nc{\aX}{{\widetilde X}}
\nc{\norm}{{\mathfrak n}}
\nc{\ndis}{\ensuremath{\texttt{dis}}}
\nc{\jj}{\ensuremath{\mathtt{j}}}
\nc{\Xthr}{X^{ThR}}
\nc\vv{\mathbf{v}}
\nc\ww{\mathbf{w}}
\nc\xx{x}
\begin{document}

{	\title{Charges of long random states \\
		of the Heisenberg spin-1/2 chain model}
	
	\author{Jacek Pawe{\l}czyk\footnote{\texttt{jacek.pawelczyk@fuw.edu.pl}}\\ \small
		Institute for Theoretical Physics,\small
		University of Warsaw\\ \small
		Pasteura 5, Warsaw, Poland
	}
	
	\date{\today~~~~}
	
	\maketitle
	
	\abstract{We conjecture a formula which expresses charges of
	infinitely long states of the Heisenberg spin-1/2 chain model. Several arguments are provided which support the proposal.
	}
	}
	\newpage

\tableofcontents

\section{Introduction}


Integrable spin chains are very lively developing realm of theoretical physics
\cite{integrable,integrable-2}. They are characterized by an infinite set of conserved charges all of them encoded in the transfer matrix $T$. 
 The Heisenberg su(2) spin-1/2 chain is one of the simplest and most studies integrable models
The standard $T$ is a trace of the monodromy matrix for which the auxiliary space is in the fundamental representation of the symmetry group of the model: in our notation it is $\jj=1$.
Recently it has been shown that the models have additional conserved charges originating from transfer matrices 
$T_\jj$ for which the auxiliary space is in a higher spin representation\footnote{Here $\jj=2 s\in \NN$ where $s$ is spin.} of su(2) \cite{Ilievski:2015nca}.  
Their existence requires  spin chains to be infinitely long. 
The higher spin charges are not strictly speaking local but only so-called quasi-local operators. 
Soon after discovery  they appeared to be necessary in description of steady-state averages after quantum quenches \cite{gge-ext}.  The proper framework in  this case involves so-called 
 Generalized Gibbs Ensemble \cite{gge} which must include all the conserved charges and the corresponding chemical potentials.

In this paper 
we are going to discuss charges $X_\jj$ of states of the form $\Psi=(\psi)^{\otimes N/M}$, where $\psi$ has length $M$ of the infinite ($N\to\infty$) periodic  Heisenberg model.
The work is centered around  a conjecture which, roughly 
speaking, says that for most of the very long $\psi$'s  charges $X_\jj$ are
well approximated in the physical strip (PS) of the complex spectral parameter $\mu$ by very simple formula
\beq\label{hyp}
X_\jj(\mu)=\frac1{4\pi}\frac{{\jj}}{\mu^2+\frac{1}{4} (\jj+1)^2}\ .
\eeq
When the length of $\psi$ goes to infinity we expect that \refeq{hyp} is exact.
Notice that \refeq{hyp} depends only on \jj.  
We shall be more specific about the precise meaning of the hypothesis in Sec.\ref{sec:approx}. 

We support \refeq{hyp} providing several arguments. First of all we derive analytically, under certain assumptions,  the formula in the case $\jj=1$. 
Next,  we do certain large $\mu$ expansion which, in fact, coincides with \refeq{hyp} for general $\jj$. and do statistical analysis of vast numerical data obtained mostly for $\jj=1$ and $\jj=2$. Finally we  formulate the conjecture and then show that it is in agreement with infinite temperature average in Gibbs ensemble.

The hypothesis claims enormous simplification of charges for some  states. This simplicity is quite astonishing in 
view of known complexity of exact results. Sizes 
of expressions on $X_\jj$ grows rapidly
 with the state's length and $\jj$:  
 few examples will be given in Sec.\ref{app:pic}.  

The paper is organized as follows. In the next section we introduce definition of charges and present explicit expressions helpful in calculations. 
We also discuss some
 features of the exact formulae on $X_\jj$ (Sec.\ref{sec:anal}).
  Sec.\ref{sec:approx} contains main results of the paper leading to our hypothesis. Thus we first discuss a large $M$ limit of just $X_1$ for which we can do analytic calculations. Next we do large $\mu$ approximation.
 Finally we compare numerically \refeq{hyp} and the exact results on charges of randomly chosen and quite long (up to length 200) states. The main body of paper ends with Conclusions.
Several appendices contains details on notation and  technical aspects of the results. 
\section{Charges}
In this section we recall definitions of 
 charges of the spin  chain state $\Psi$ \cite{Ilievski:2015nca}. The presentation culminates with the expression on $X_\jj(\mu)$ which will be used in the next sections.

Conserved quantities of the integrable su(2) Heisenberg spin chain model of the length $N$ are given by the expectation value of the transfer matrix:
 $$
 T_{(0)}^\Psi(\mu)=\langle\Psi|T_{(0)}(\mu)|\Psi\rangle=\langle\Psi|\tr_0 (L_{0N}L_{0(N-1)}\dots L_{0k}\dots L_{01})(\mu)|\Psi\rangle,
 $$
 where $L_{0k}$ is the Lax operator, "$0$" denotes an auxiliary space, $k$ the k-th node of the chain
 and $\Psi$ is a given spin chain state belonging to  quantum space 
 ${\cal H}= ({\cal V})^{\otimes N}$, $\dim {\cal V}=2$.
 Then $X_{(0)}(\mu)\sim \p_\mu T_{(0)}^+(\mu)$
\footnote{Here $f^\pm(\mu)=f(\mu\pm\frac i2)$ and we shall also use 
$f^{[\pm k]}(\mu)=f(\mu\pm k\frac i2)$. We shall often suppress any decoration of $\MM,\,\LL,\, \vv,\, \ww$ if from the context it will be clear what are $\Psi,\,\psi$ and $\jj$.}. 
Usually the  spin-$\frac12$ auxiliary space (in our notation it is $\jj=1$) is considered and then $X_{(0)}\equiv X_{\jj=1}\equiv X_1$ exist for any finite $N$. For higher spin auxiliary spaces $\jj=2,...$ the charges $X_\jj$ also exist but they are independent on $X_1$ only in $N\to \infty$ limit.
\beq\label{charge-1}
X_\jj(\mu)=\lim_{N\to\infty}\frac1{2\pi i N} \langle\Psi|\p_\mu\log\frac{T_\jj^+(\mu)}{T_0^{[\jj+1]}(\mu)}|\Psi\rangle
\eeq
 The factor $T_0^{[\jj+1]}(\mu)$ appearing in the denominator of \refeq{charge-1}  shift $X_\jj$ by the state independent function. It was introduced for convenience. 
 Strictly speaking $X_\jj(\mu)$ depends on the spectral parameter $\mu$ thus it is the generating function of the charges. In this paper we shall keep calling functions
 $X_\jj(\mu)$ charges of $\Psi$.
 
One can differentiate $\log$ producing
\beq\label{charge}
X_\jj(\mu)\ =\ \lim_{N\to\infty}\frac1{2\pi i N} \langle\Psi|\frac{ T_\jj^-(\mu)}{T_0^{[-\jj-1]}(\mu)}\p_\mu
\frac{ T^+(\mu)}{T_0^{[\jj+1]}(\mu)}|\Psi\rangle
\eeq
with the help of  so-called inversion formula, which says that
$T_\jj^-T^+_\jj=T_0^{[-\jj-1]}T_0^{[\jj+1]}$ in the $N\to\infty$ limit \cite{Ilievski:2015nca,inversion-1,inversion-2,inversion-3}.

Taking $N\to\infty$ is always a delicate matter. One must carefully define the whole procedure. Here we consider certain family of states $\Psi$ of the form
$\Psi=(\psi)^{\otimes N/M}$, where the substate $\psi$ has length $M$. 
Following \cite{double-1,double-2} we define composite two-channel Lax operator (see App.\ref{app:notation} for the notation)
\beq\label{LL}
\LL_\jj(\mu,x)=\norm(\mu,x)\,{L_\jj^-(\mu)L_\jj^+(x)}
\eeq
where $L_\jj$ denote the Lax operator in the representation $\jj$ and $\norm$ is a normalization factor originating from   $T_0^{[-\jj-1]}T_0^{[\jj+1]}$ in \refeq{charge}.
We define a monodromy operator as
\beq\label{mpsi}
\MM^\psi_\jj(\mu,x)=\langle\psi|\LL_\jj^{(M)}...\LL_\jj^{(1)}(\mu,x)|\psi\rangle
\eeq
Then
\beq\label{X-by-M}
X_\jj^\psi(\mu)=\lim_{N\to\infty}\,\frac1{2\pi i N}\ \tr_{V_\jj\otimes V_\jj}
[\p_x(\MM^\psi_\jj(\mu,x))^{N/M}]\vert_{x=\mu}
\eeq
The operator $\MM_\jj^\psi(\mu,x)$  has generically one eigenvalue $\la$
which tends to 1 when $x\to \mu$ \cite{double-2}.
\beq\label{sch}
\MM_\jj^\psi(\mu,x)\vv_\jj(\mu)=(1+(x-\mu)\, \d_\jj\,)\vv_\jj(\mu)+\cO((x-\mu)^2)
\eeq
The unit eigenstate and it eigenvalue dominate the trace $\tr$ in \refeq{X-by-M} 
\beq\label{ch1}
X_\jj^\psi(\mu)=\lim\frac1{2\pi i N}(\p_x (1+(x-\mu) \d_\jj)^{N/M}|_{x=\mu})=\frac1{2\pi i M}\d_\jj
\eeq
It follows from \refeq{sch} that we can find left and right unit eigenvalues of $\MM^\psi\equiv\MM^\psi(\mu,\mu)$
\beq\label{eigenv}
\MM_\jj^\psi\, \vv_\jj=\vv_\jj\ ,\; \ww_\jj^\dagger\,\MM_\jj^\psi =\ww_\jj^\dagger
\eeq
By standard quantum mechanical perturbative calculations one obtains
\beq\label{ch2}
\d_\jj=\frac{\ww_\jj^\dagger\,\p\MM_\jj^\psi\, \vv_\jj}{\ww_\jj^\dagger\, \vv_\jj}|_{x=\mu}
\eeq
where $\p\MM^\psi_\jj\equiv\p_\xx\MM^\psi_\jj(\mu,\xx)|_{\xx=\mu}$.
The above expression is equivalent to what was derived in \cite{double-2}\footnote{One can easily show that
$\vv_\jj\,\ww_\jj^\dagger\sim  \mbox{Adj}(\MM-1)$.}
It appears to be very handy for various types of calculations presented in some details in Appendices. It is specially useful for efficient numerical calculations when  $\psi$'s are, what we shall call, simple substates. In that case $\psi$ is just single sequence of spins up and down represented by $1$ and $2$. The reason for this simplification follows from triviality of the right unit eigenvector $\ww$. For more details we send the reader to App.\ref{app:notation}.

\subsection{Charges $X_{\jj}$'s and their analytic structure}
\label{sec:anal}
The charges $X_\jj(\mu)$ have very reach analytic structure on complex $\mu$-plane. 
For simple $\psi$ they are rational functions with poles and zeros which number grows 
rapidly with their length and the representation index $\jj$.  
Higher spin charges of long $\psi$ have mammoth sizes, thus they are completely impractical. To give the reader a flavor how the exact expressions may look like we display an example ($M=7,\, \psi=\{1, 1, 1, 1, 2, 1, 2\}$) which still fit in the paper: see App.\ref{app:pic}. For larger length $\psi$ the formulae would occupy several pages e.g. for $M$=40 and typical  $\psi$ the charge $X_3(\mu)$ denominator is a polynomial of the degree 234 with integer coefficients containing 80 digits. 

\begin{figure}[!]
\centering
\includegraphics[width=5.5cm]{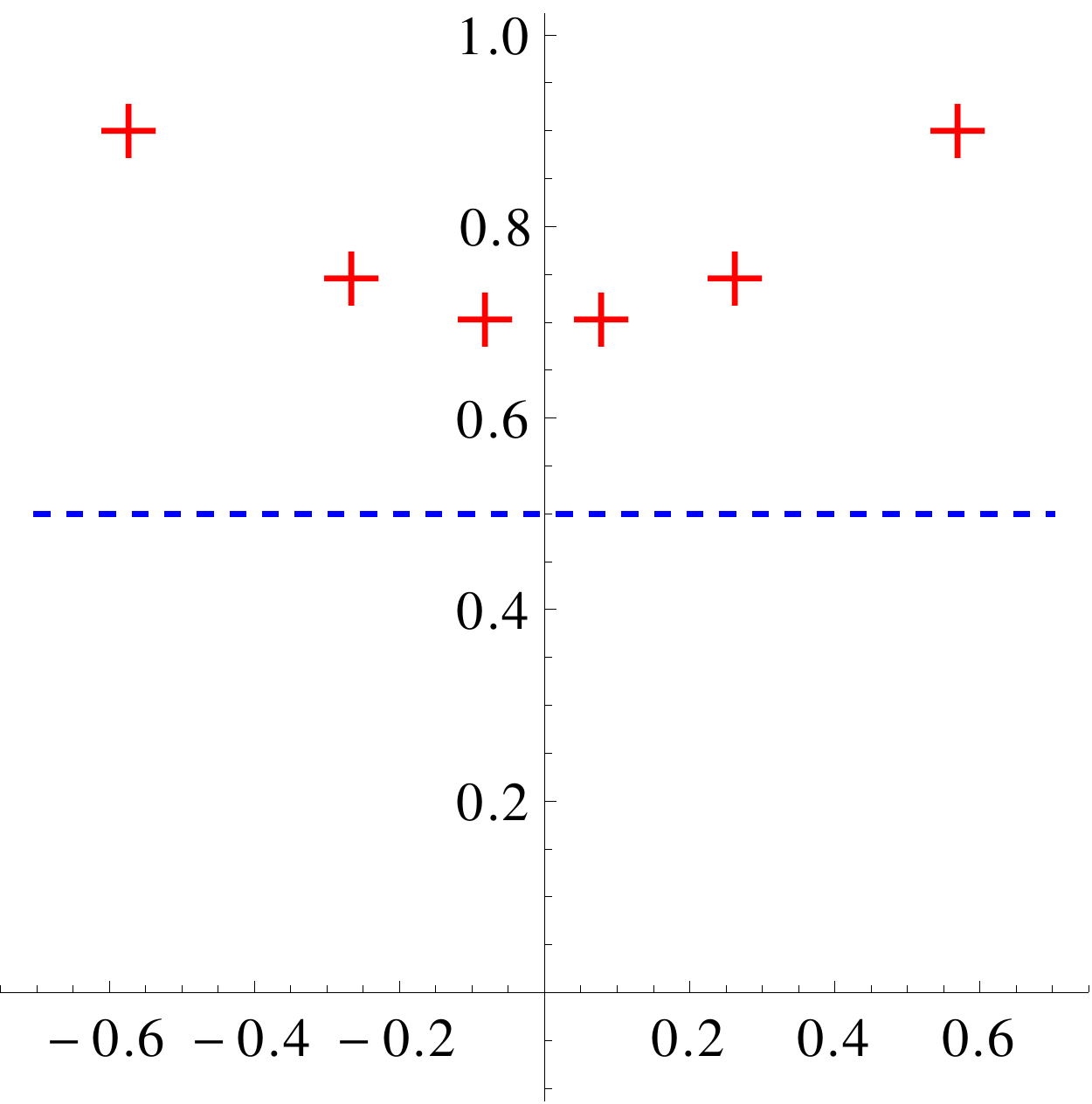}\hskip1.5cm
\includegraphics[width=5.5cm]{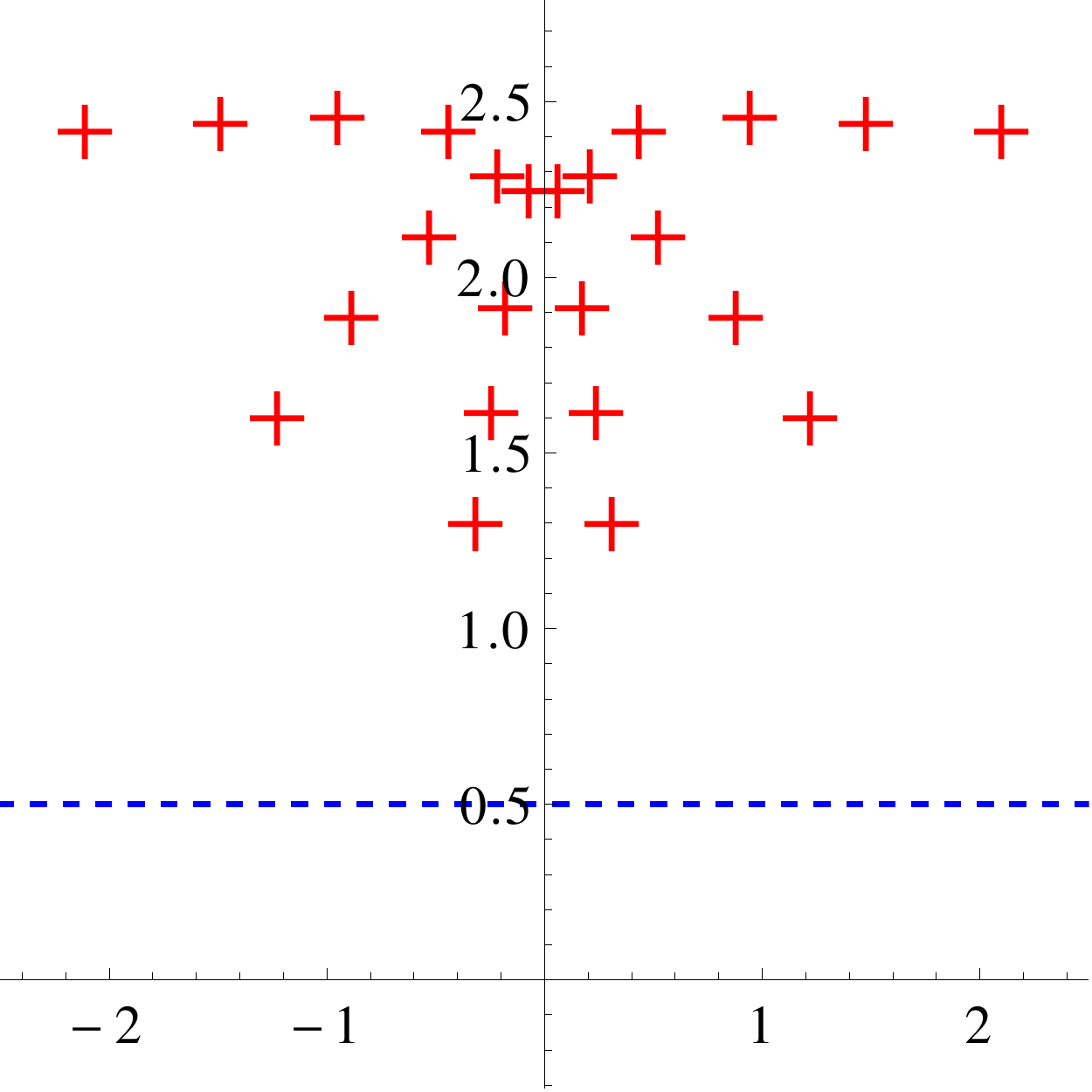}
\caption{Positions of poles of $X_1(\mu),\ X_4(\mu)$ for $\psi=\{1, 1, 1, 1, 2, 1, 2\}$ on the complex half-plane ($\mu,\ \Im(\mu)\geq 0$). The explicit expression for $X_4$ would occupy too much space here, thus we present $X_1$ only: $X_1(\mu)=\frac{1}{7\pi}\times\frac{10 \mu^{10}+25 \mu^8+34 \mu^6+26 \mu^4+11 \mu^2+2}
{7\mu^{12}+21 \mu^{10}+35 \mu^8+35 \mu^6+21 \mu^4+7 \mu^2+1}.$
}
\label{fig:X-ch-7}
\end{figure}

It is much easier to see structure of charges displaying their poles 
 on the complex $\mu\geq 0$ half-plane\footnote{For simple $\psi$'s charges $X_\jj$ are even and real functions of $\mu$.}.
The examples are shown of Fig.\ref{fig:X-ch-7}. 
In spite of complexity $X_\jj(\mu)$'s possess several general properties which can be spelled out.
\ben
\- All poles in $X_\jj(\mu)$ appears beyond the so-called Physical Strip (PS) which is:\\
PS $=\{\mu\in\CC; |\Im(\mu)|<1/2\} $. The poles we interpret as bound states of auxiliary spins in the background of fixed $\Psi$ (see also \cite{babelon}).

On the technical level poles originates from zeros of $\ww^+\vv$ and corresponds to those $\mu$'s for which $\MM^\psi_\jj$ contains nontrivial 2-dim Jordan block with eigenvalue 1.  
 The proof of this statement is 
given in App.\ref{app:X1} for the simplest case of $X_1$ only. 
For the higher charges we checked  that fact numerically only.
\- Poles of $X_1$ lay on the hyperbola $\re(\mu^2)=-1/2$ which represents the relativistic dependence of the real and imaginary part on $\mu$ (App.\ref{app:X1}).
 Poles of the higher charges seems to align certain curves too, but their nature is more complicated (see Fig.\refeq{fig:X-40-23} in App.\ref{app:pic}).
\- Most of the poles and zeros of $X_\jj$'s are very close to each other what means that their contribution to charges inside PS is very small. This suggest big redundancy of information contained in exact expressions. 
\een
\section{The conjecture}
\label{sec:approx}

From the analysis presented in the previous sections it is clear that the exact structure of the charges is very complicated. For long chains one may doubt if exact expressions on $X_\jj$ (if known) would be of any practical use \footnote{We leave aside experimental problems related to the ability to control initial condition of spins in chain i.e. the state $\psi$.}. 
Thus a formula that would well approximate charges in PS might be very useful.
In this section we shall make proposal which seems to do the job for very long random and simple $\psi$. First we present arguments which will justify our final statement of Sec.\ref{sec:hyp}.

\subsection{$X_1$ in $M\to\infty$ limit}
\label{sec:X1-int}
One may wonder what is the distribution of poles thus  the density of charges
on the  complex $\mu$-plane in large $M$ limit.
We are not going to consider here the most general case of arbitrary $X_\jj$.
Quick look at the distribution of poles suggests that the problem might be very hard. But for $X_1$ thanks to the results of App.\ref{app:X1} we can present calculations which lead to conceivable picture of the limit. The procedure we propose is a direct analog of the thermodynamic limit \cite{limth-1,limth-2}.

\begin{figure}[h] 
	\centering
	\includegraphics[width=6.5cm]{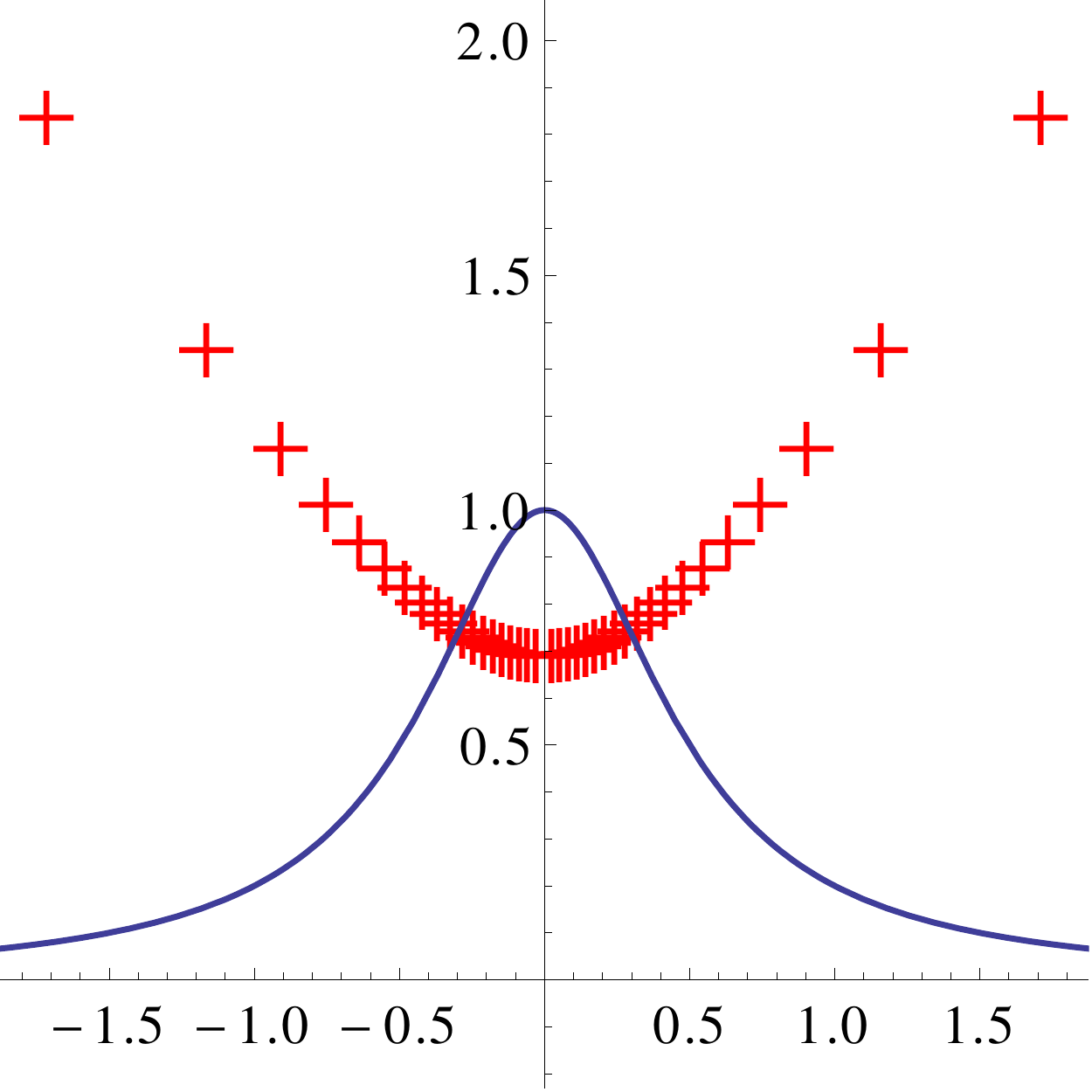}\hskip1cm
	\includegraphics[width=7.5cm]{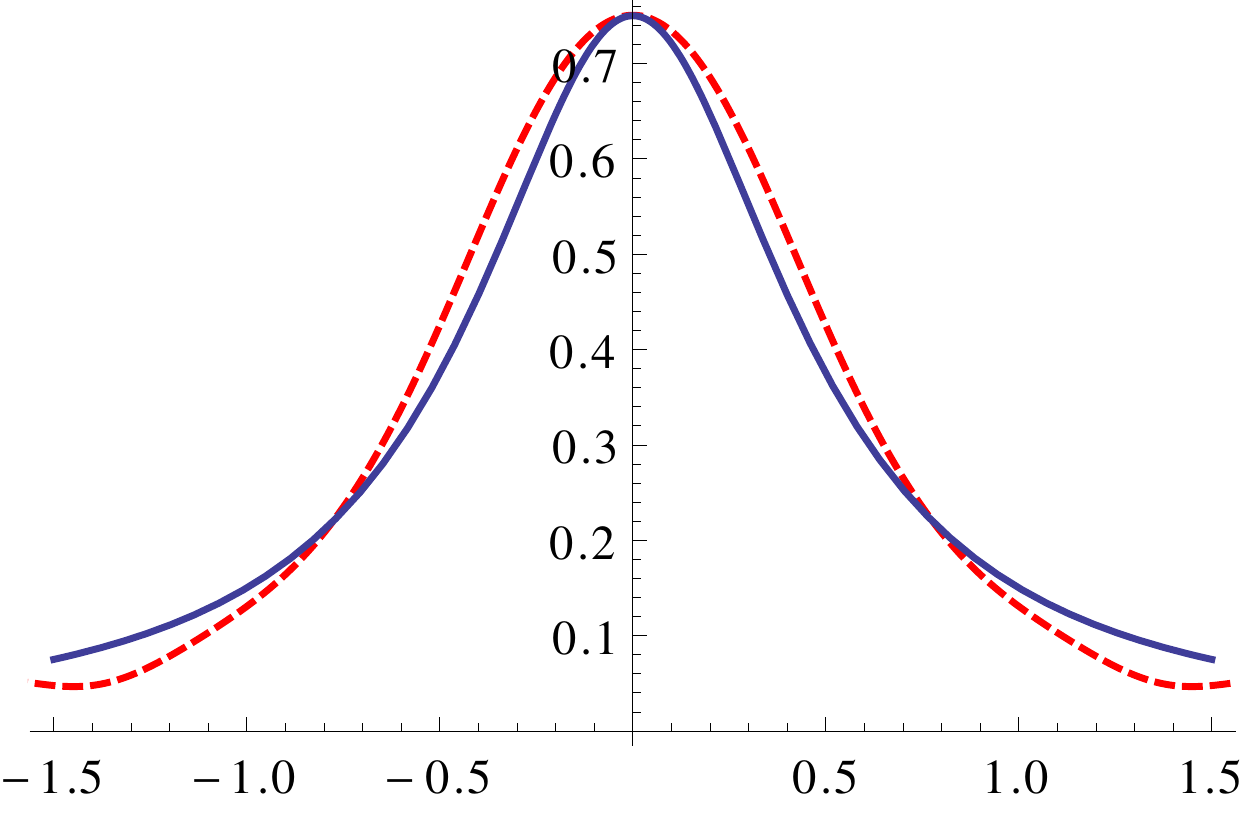}
	\caption{(a) Poles of $X_1$ for {\footnotesize  $\psi=\{1,2,1,1,1,2,2,1,1,1,1,2,1,2,1,1,2,1,1,1,1,1,2,1,2,1,2,1,
			2,2,1,2,\ $ $\ 2,2,1,2,1,1,2,1\}$} marked by red crosses and the density $1/(4a^2+1)$ of \refeq{X1-int} as a function of $a=\Re(\mu)$ (blue line). 
		(b) Pole density of $X_1$ drawn by \texttt{SmoothHistogram} of \texttt{Mathematica} as a function of
		$a=\Re(\mu)$  (red dashed line) and appropriately scaled density $1/(4a^2+1)$  (blue line).}
	\label{fig:X-40-1}
\end{figure}

First of all we decompose the formula on $X_1$ as
\beq\label{X1-sum}
X_1=\frac1{2\pi}\sum_{n}\frac{c_n}{\mu-\mu_n}
\eeq
where $\mu_n$ are  positions of poles and $c_n\in \CC$. From App.\ref{app:X1} we have 
$$
\mu^2_n=i\, y_n-1/2,\quad \frac{i\,y_n-1/2}{i\,y_n+1/2}=e^{2i\pi k_n/M},\quad k_n=1,...M-1,\; y_n\in\RR
$$
One must remember that for non-generic substates not all $k_n$ correspond to poles: there are holes in the distribution $k_n$ i.e. there are less poles then $M-1$.
 At large $M$ and generic random $\psi$ we expect that there are no holes i.e. there is one-to-one $k_n\leftrightarrow n$ linear relation thus we shall set $k_n=n$. Fig.\ref{fig:X-40-1} shows  poles for $M=40$ state. 
Notice that density of a charge is not given uniquely by distribution of poles. This would hold only if all residues $c_n$ were equal what we are going to assume from now on. Thus we set $c_n=c$. Denoting the continuous variable
as $\a$ i.e. $n/M\to \a$ in $M\to\infty$ limit we rewrite \eqref{X1-sum} as
\beq
\Xthr_1(\mu)=\frac c{2\pi}\int d\a\,\frac{1}{\mu-\mu(\a)}
\eeq
where
\beq
\mu^2(\a)=i\,y(\a)-1/2,\quad \frac{i\,y(\a)-1/2}{i\,y(\a)+1/2}=e^{2i\pi \a},\quad y\in \RR
\eeq
Changing integration variable to $a=\Re(\mu(\a))$ we get: \\
\beq\label{X1-int}
\Xthr_1(\mu)=\frac c{2\pi^2} \int_{-\infty}^{\infty}\frac{da}{4 a^2+1}\ \frac1{a^2+(\mu -a)^2+\frac{1}{2}}= \frac {c}{4\pi(\mu^2+1)}
\eeq
The above constant $c$ has been redefined to include numerical 
factors appearing in the course of calculations.
 The representation \eqref{X1-int} is properly
defined for $\mu\in \RR$ but can be analytically extended to the whole 
complex plane.
The final value of $c$ can be fixed by comparing with large $\mu$ result of Sec.\ref{sec:large-mu} and App.\ref{app:large-mu}  yielding $c=1$.

From \refeq{X1-int} the density of charge in the variable $a$ can be read to be
$1/(4a^2+1)$. The letter is included in Fig.\ref{fig:X-40-1} and it nicely 
fits the density of poles on the hyperbola. 

Ww want to stress that $c_n\to c(\a)=c=const$ can not hold for general $\psi$ e.g.  for  $\psi$'s which are of the form $\psi=(\psi')^{M/M'}$, 
where the length of $\psi'$ is $M'$. Then $X_1^{\psi}=X_1^{\psi'}$
thus it does not depend on $M$ at all. In the extreme case $\psi'=\{1,2\}$ (N\'eel state) 
the whole density $c(\a)$ is localized at two points $\mu=\pm i/\sqrt{2}$. 
On the other hand we expect that for most of the  
long random $\psi$'s, $c(\a)=c=const$ is good approximation.
 This point will be under thorough scrutiny in Sec.\ref{sec:num}. 
\subsection{Large $\mu$ expansion}
\label{sec:large-mu}
In this section we shall discuss 
certain large $\mu$ approximation of the exact expression on $X_\jj$. 
The procedure we propose is a hybrid: we do large $\mu$ expansion of the Lax operators and the vector $\vv$ but keep intact the normalization factor $\norm$. This is well motivated by the previous derivation of Eq.\refeq{X1-int} where $\norm$ appears naturally from continuous distribution of the charge density localized along a hyperbola. 
Detailed derivation is presented in App.\ref{app:large-mu}. 
The obtained result is
\beq\label{approx}
\tilde{X}_\jj(\mu)\approx\frac1{2\pi}\frac1{\mu^2+\frac{1}{4} (\jj+1)^2}\left(\frac{\jj}{2}-\frac{\jj (1-r) \left(r^{\jj+1}+1\right)-2 r \left(1-r^\jj\right)}{2
   (r+1) \left(1-r^{\jj+1}\right)}\right)
\eeq
where $r=n_2/n_1$ and $n_1,\ n_2$ denote numbers of spins up and down in $\psi$, respectively. The formula is $r\to 1/r$ invariant.
Few remarks are necessary at this point. The singularities at $\mu=\pm i 
\frac{\jj+1}2$ come from normalization factor $\norm(\mu)$. In the approximation made the denominator of
 \ref{ch2} i.e.
 $\ww^+\vv$ is $\mu$  independent contrary to exact results on charges.
 Recall that zeros of $\ww^+\vv$ give spectra of bound states. These we do not  expect to appear in $\mu\to \infty$ limit, at least at the leading order of the expansion. Thus $\ww^+\vv=const$ is physically well motivated.

The r.h.s. of \refeq{approx} has the following expansion for small 
$\ep\equiv 1-r$:
\beq\label{approx-ep}
\tilde{X}_\jj(\mu)\approx\frac1{4\pi}\frac{\jj}{\mu^2+\frac{1}{4} (\jj+1)^2}\left(1+\frac1{12}\ep^2\right)+\cO(\ep^3).
\eeq
For random states of the length $M$ the average deviation of $|\frac{n_1-n_2}{n_1+n_2}|=|(1-r)/(1+r)|\sim {\sqrt{M}}/M\to 0$ when $M\to\infty$ thus $\ep=0$ for most of the random long $\psi$'s.

Several pictures comparing $X_\jj$ and $\tilde{X}_\jj$ are given in App.\ref{app:pic} (Fig.\ref{fig:ndis}). From there we see that
 \refeq{approx} works quite well even for 
 relatively short $\psi$'s. Moreover the higher representation \jj{} the better are approximations. But we need more quantitative checks.
  The next section is devoted to a simple statistical analysis of estimates provided by \refeq{approx}.
\subsection{Statistics}
\label{sec:num}
It is interesting to check how well \refeq{approx} estimates the exact expression. Previously given arguments for $X_1$ gives hope that the proposed formula is, in a sense, exact in the $M\to\infty$ limit. Thus \refeq{approx} for $\jj=1$ should be good approximation even for large but finite $M$. The situation is less clear for $\tilde{X}_\jj$ ($\jj>1$) where we do not have similar analytical arguments for
higher charges thus we are forced to rely on statistical analysis only.
 Moreover
due to length of exact formulae we have been unable to go too far with value of $M$  and $\jj$. 

\begin{figure}[!]
\centering
\includegraphics[width=10cm]{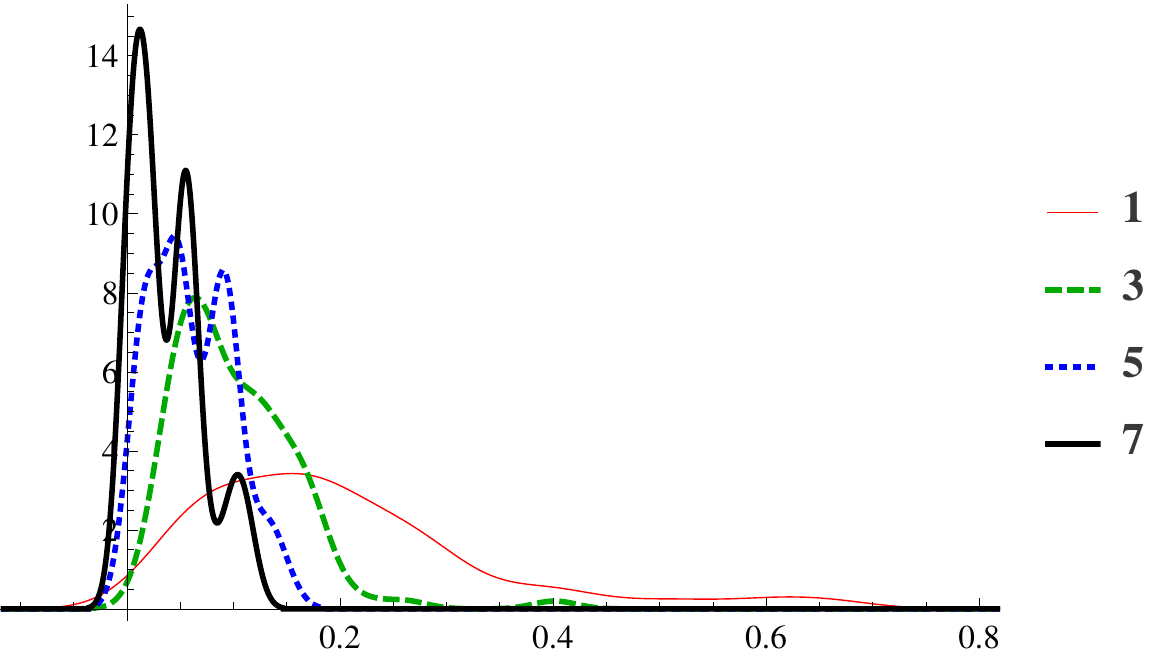}
\caption{Histogram of deviation $\ndis_\jj$ for $\jj=1,3,5,7$ for
random chains  $M=20$.}\label{fig:hist_j}
\end{figure}

\begin{figure}[h]
\centering
\includegraphics[width=10cm]{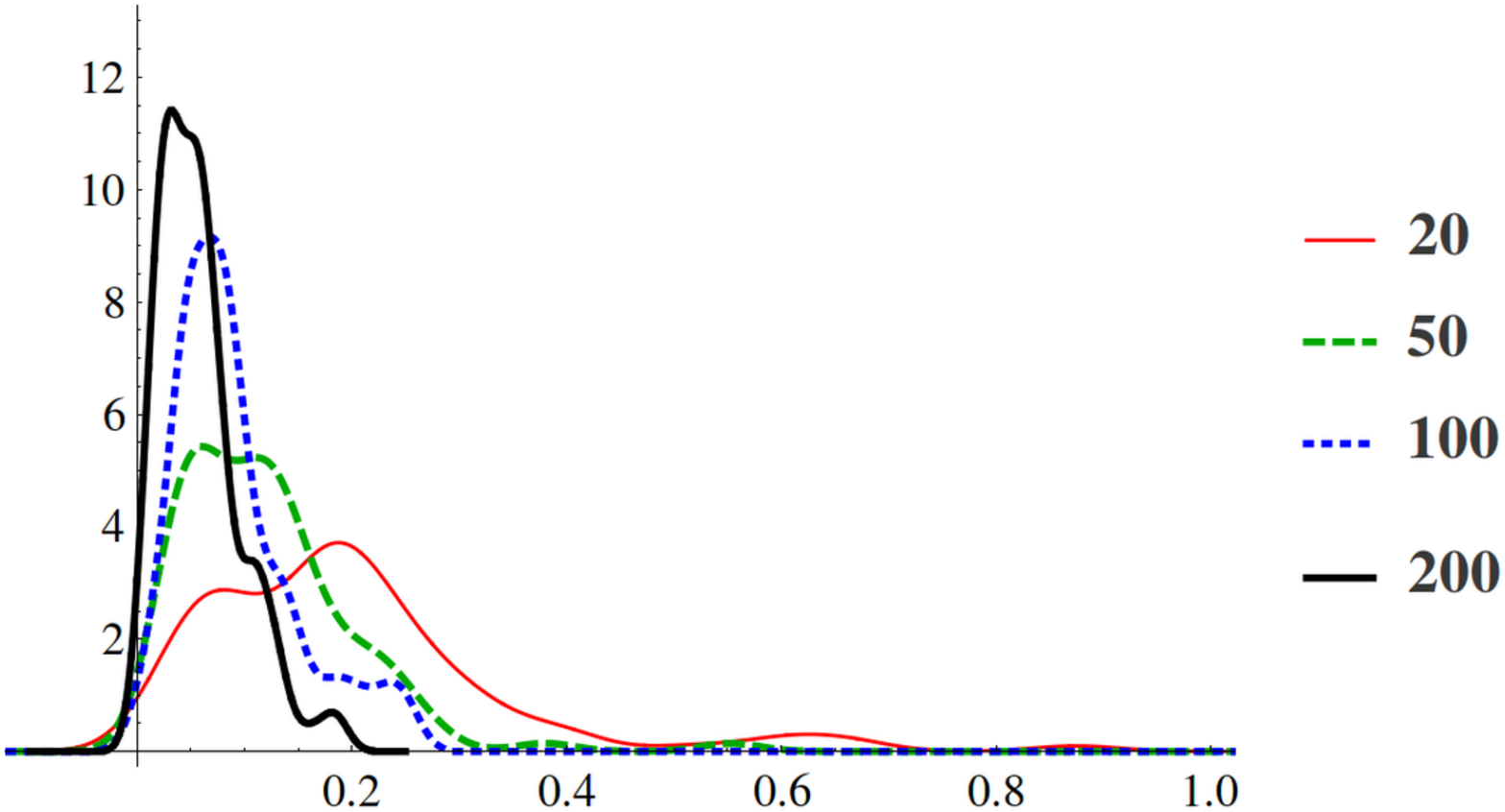}
\vskip-1cm
\caption{Histogram of deviation $\ndis_1$ for $M=20,50,100,200$}\label{fig:hist:o}
\end{figure}

Hereafter we shall compare values of $X$'s and $\tilde X$'s on the real 
line $\mu\in \RR$. 
As a measure of deviation between $X_\jj$ and $\tilde X_\jj$ we have chosen
\beq\label{ndis}
\ndis_\jj=\mbox{sup}_{\mu\in[-10,10]}\left|\frac{X_\jj(\mu)-{\tilde X}_\jj(\mu)}{X_\jj(\mu)}\right|
\eeq
which will be calculated  for the following cases:
\ben
 \item[(a)] fixed 100 random $\psi$'s of the length $M=20$ for different representation index $\jj=1,3,5,7$
 \item[(b)] $\ndis_1$ calculated for 100 random $\psi$'s of the lengths $M=20,\ 50,\ 100,\ 200$.
 \item[(c)] $\ndis_2$ calculated for 100 random $\psi$'s of the lengths $M=20,\ 50,\ 100$.
 \een
The obtained data were plotted on histograms Fig.\ref{fig:hist_j} for the case (a) and Fig.\ref{fig:hist:o} for the case (b)\footnote{All calculations have been done by \texttt{Mathematica}.  We have used \texttt{RandomInteger[{1,2},20]} as generator of random $\psi$ of $M=20$. Negative values of $\ndis_\jj$ in Figs.\ref{fig:hist_j} and \ref{fig:hist:o} follows from  interpolation done by \texttt{SmoothHistogram} function.}. 
For the case (c) i.e. $X_2$, the histogram appears to be very similar to (b) hence it is not displayed here.

It is clear that the bigger $M$ the relative difference between $X_\jj$ and $\tilde{X}_\jj$
is smaller. For $\jj=1$ and $M=50$ the deviation $\ndis_1$ for random substates peeks about 0.1 while for $M=200$ it is only 0.05. Similar tendency is seen for $\jj=2$ but we had  poorer statistics in this case. 
Moreover Fig.\ref{fig:hist_j} suggests that statistically the formula works better if 
the representation $\jj$ is higher although we did not do enough numerics to 
make any convincing claim to what extend $\tilde{X}_\jj$ works better for 
e.g. $\jj=3$ compared to $\jj=1$.

It is of primer necessity to increase  amount of numerical data to support \refeq{approx} and our main conjecture discussed in the next paragraph.

\subsection{The conjecture} 
\label{sec:hyp}
In this section we shall spell out our main hypothesis and clarify some of vague statements appearing in the paper. 
Our claims are based on arguments given in the previous subsections. Moreover we present new reasons which let us 
 extend the conjecture to non-simple $\psi$'s. 

Substates $\psi$ of the previous section  have been chosen randomly. The random choice include those 
$\psi$'s which  charges are far from being close to \refeq{approx}.
These we call non-generic. For example: $\psi=(\psi')^{\otimes M/M'}$
($\psi'$ has length $M'$, $M'$ is a nontrivial divisor of $M$) are non-generic: 
 $X_\jj^{\psi}=X_\jj^{\psi'}$ for any $M$. The important fact (supported by numerics of the previous subsection) is that for
large $M$
probability that random $\psi$ is non-generic is close to zero. In this sense the conjecture is formulated for most of simple $\psi$'s.


The space of states of the model is very reach but 
up to this point we have been solely working with simple $\psi$'s in the form of one sequence of spins up and down.  These are rare in the space of all states. 
The most general $\psi$'s are of the form
 \beq
 \psi=\sum_n \a_n\psi_n\, ,\quad \a_n\in\CC
 \eeq
where now $\psi_n$'s are all different and simple. Hence we need to calculate 
\beq\label{psimn}
\langle\psi_m|\LL^{(M)}...\LL^{(1)}(\mu,x)|\psi_n\rangle,\quad m\neq n
\eeq
for all $m,n$. 
 The claim is that if both $\psi_m$ and $\psi_n$, $m\neq n$ are random then the above expression vanish in $M\to\infty$ limit.
The crucial point is that \refeq{psimn} always contains  off-diagonal terms of $\LL$ i.e. $\LL^1_2$
 and $\LL^2_1$ which  number grows to infinity when $M\to\infty$.
Inspection of
\refeq{lls} reveals that $\LL^1_2(\mu,\xx)$ and $\LL^2_1(\mu,\xx)$ are contracting operators 
 i.e. 
$\vert|\LL^1_2(\mu,\xx)\cdot v\vert|\leq p\vert|v\vert|\; v\neq 0,\, v\in\CC^{2\jj+2}$ and $q\in[0,1)$
  for any   representation \jj{} and $\mu,\, \xx\in\RR\backslash\{0\}$  (the same holds for $\LL^2_1$). Indeed, e.g. for $\jj=1$  we have
 \beq
\frac{\vert|\LL^1_2(\mu,\xx)\cdot v\vert|^2}{\vert|v\vert|^2}=\frac{\vert|(-i+\mu)v_2+(i+\xx)v_3\vert|^2+(\mu^2+\xx^2)\vert|v_4\vert|^2}{(1+\mu^2)(1+\xx^2)}\leq 1
 \eeq
where the equality can hold only for $\mu=\xx=0$. For higher $\jj$ the bound  $p$
 is smaller then 1 e.g. for $\jj=2$ it is $p=8/9$ for all $\mu,\,\xx\in R$.
 Infinite product of  contracting operators and bounded by 1 operators $\LL^1_1,\,\LL^2_2$  yields zero. Assuming that analyticity in $\mu,\, \xx\in$ PS is preserved by the limiting procedure we infer that \refeq{psimn} vanishes. Thus if the hypothesis is true for simple $\psi$ it is true for all long, random $\psi$. 

 {\vskip.2cm
 \paragraph{Conjecture.} For almost all states of the form $\Psi=(\psi)^{\otimes N/M}$ ($N$ is divisible by $M$) where $\psi$ is 
 a random substate of the length $M$ 
  the charges \refeq{charge} in  the limit  $M\to\infty$ are given by:
 \beq\label{approx-0}
 \lim_{M\to\infty} X_\jj(\mu)\equiv \Xthr_\jj(\mu)=\frac1{4\pi}\frac{\jj}{\mu^2+\frac{1}{4} (\jj+1)^2}
 \eeq
 \vskip.2cm}

\subsection{$T\to\infty$ average}
The conjecture might be very hard to prove by direct means as it has been discussed in previous sections. But if correct it has direct consequences which can be easily checked. 
Here we shall calculate the average of the charges over infinite temperature Gibbs ensemble for infinitely long spin chain
and show that it is equal to the r.h.s of \refeq{approx-0} 
\footnote{The calculations has been suggested to the author by 
Bal\'azs Pozgay, Jacopo de Nardis, Enej Ilievsky and Mi{\l}osz Panfil.}. 
This should be expected if states of charge \refeq{approx-0} dominates the ensemble.

There is another arguments in favour of the relation to the above $T\to\infty$.
Notice that charges determine equilibrium densities through string-charge relations of \cite{string-charge}. 
 \beqa
  \rho_\jj&=&X_\jj^++X_\jj^--X_{\jj-1}-X_{\jj+1}\\
  \bar{\rho}_\jj&=&\frac1{2\pi}\frac{4 \jj }{\jj^2+4 \mu ^2}-X_\jj^+-X_\jj^-
 \eeqa
For \refeq{approx-0} we get:
  \beqa
  \rho_\jj(\mu)&=&\frac1{2\pi}\frac8{(4\mu^2+\jj^2)(4\mu^2+(\jj+2)^2)}\\
  \bar{\rho}_\jj(\mu)&=&\frac1{2\pi}\frac{8\jj(\jj+2)}{(4\mu^2+\jj^2)(4\mu^2+(\jj+2)^2)}
  \eeqa
Thus the ratio of holes to particle densities is determined to be constant
  depending only on \jj : $\eta_\jj=\jj(\jj+2)$ . The latter respects Y system \cite{y-system-1,y-system-2,y-system-3,y-system-4}
  \beq
  \eta_\jj^+\eta_\jj^-=(1+\eta_{\jj+1})(1+\eta_{\jj-1})
  \eeq
 which is equivalent to TBA in some cases \cite{limth-2, y-mod}. 
 Here it is 
 $T\to\infty$ limit of TBA (see \cite{limth-2}).

The average is defined as
 \beq\label{average}
 \langle X_\jj \rangle =\lim_{N\to\infty}\frac1{2\pi i N}\tr_{V_\jj\ox V_\jj}
 \left. \p_x\left( \half\tr(\LL_\jj(\mu,x))\right)^N\right|_{x=\mu}
 \eeq
 where the inner trace is over single node quantum space. Explicitly 
 \beqa
 \frac12\tr(\LL_\jj(\mu,x)) 
 =\frac{\norm(\mu,x)}{4} \left((2 \mu -i) (2 x +i)
 -2\, C_2
 \right)
 \eeqa
 where $C_2=\left(\hat{s}^-\otimes \hat{s}^++\hat{s}^+\otimes
    \hat{s}^-+2 \hat{s}_z\otimes \hat{s}_z\right)$ is a Casimir acting on $V_\jj\ox V_\jj=\oplus_{r=0}^{2\jj} V_r$.
 Eigenvalues of the $\LL_j$ for the $r$-representation $V_r$ are:
 \beq
 \la_r=\frac{(\jj+1)^2-r (\frac{r}2+1)+2 i (\mu -x )+4 \mu  x}{(2 \mu -i 
 (\jj+1)) (2 x +i (\jj+1))},\quad r=0,...2\jj.
 \eeq
 Only $r=0$ term survives the limit $N\to\infty$ in \refeq{average} yielding:
 \beq
 \frac1{2\pi i N } \left. \p_x(\la_0^N)\right|_{x=\mu} \to  \frac1{\pi}
 \frac{\jj}{(\jj+1)^2+4\mu^2}
 \eeq
 what is the expected result.
\section{Conclusions}
In this paper we conjecture a formula $\Xthr_\jj$ expressing conserved charges
 of very long random states $\Psi=(\psi)^{\otimes N/M}$ of the Heisenberg spin chain. If the length $M$ of the substate $\psi$ goes to infinity the claim 
 is that the formula is exact. Otherwise it provides a good approximation of
  a very complicated exact expression. In the case $\jj=1$ we have been able to derive $\Xthr_1$ in spirit of the standard thermodynamic limit. Unfortunately we do not have such arguments for bigger \jj.
The very striking feature of the formula is its simplicity. 
If our claim is correct this suggest existence of relatively simple analytical arguments supporting it.

We have checked   numerically  for $M$ ranging up to 200  but for relatively low  representations $\jj=1,\,2$ that the longer are $\psi$'s the conjectured formula is closer to the exact one. 
Due to lack of analytic proof it would be useful  to increase 
 amount of numerical data. 

On the way to the main result  we have also  obtained  leading terms of  a large spectral parameter expansion  of charges.  It would be interesting to investigate if one can calculate  next to leading terms  or maybe even formulate  consistent perturbative approach.  The delicate point is that such an expansion should be
 regular for all $\mu\in PS$. 
 
Finally we must mention that as a consequence of the conjecture the infinite temperature limit of the average of the charges are given exactly by \refeq{approx-0}. This strengthen our believe that the conjecture is correct. 

 \vskip2cm 
  
\paragraph{Acknowledgments}

We would like to thank M.Panfil and  Jacopo De Nardis for many valuable and inspiring discussions and to Bal\'azs Pozgay, Jacopo de Nardis, Enej Ilievsky for a fruitful exchange of letters.
\appendix
\section{Appendices}

\subsection{Basic notation}
\label{app:notation}
Although the formula (\ref{ch2}) is very explicit in practice 
higher spin charges are  difficult to calculate for general $\psi$. Things are easier when one limits considerations to simple substates being one single chain of spins up and down e.g $\psi=\{1,2,1,2,2,1,2\}$ where numbers 1,2 represent spins up and down respectively. For this state $\MM_\jj^\psi(u,x)=\prod_{i=1}^M({\LL_\jj}(\mu,\xx))^{\psi(i)}_{\psi(i)}$.
where $i$ indicates the node number and $\psi(i)=1,2$. Thus $({\LL_\jj}(\mu,\xx))^{\psi(i)}_{\psi(i)}$ are\footnote{We follow conventions of \cite{string-charge}. }:
\beqa\label{lls}
(\LL_\jj(\mu,\xx))^1_1&=&\norm\ ((\mu^-+is^z_\jj)\ox(\xx^++is^z_\jj)-s^-_\jj\ox s^+_\jj)\\ 
(\LL_\jj(\mu,\xx))^2_2&=&\norm\ ((\mu^--is^z_\jj)\ox(\xx^+-is^z_\jj)-s^+_\jj\ox s^-_\jj)\non
(\LL_\jj(\mu,\xx))^1_2&=&\norm\ (i \mu^-\otimes s_\jj^++i \,
   s_\jj^+\otimes \xx^+- s_\jj^+\otimes s_\jj^z+ s_\jj^z\otimes
   s_\jj^+)\non
(\LL_\jj(\mu,\xx))^2_1&=&\norm\ (i\mu^-\otimes s_\jj^-+is_\jj^-\otimes \xx^+ +s_\jj^-\otimes s_\jj^z-s_\jj^z\otimes
   s_\jj^-)\nn
\eeqa
where $s^a_\jj$ respects su(2) algebra in representation \jj, $\mu^\pm=\mu\pm\frac{i}{2},\ \xx^\pm=\xx\pm\frac{i}{2},\ \mu,\ \xx\in\CC$ and
\beq
\norm(\mu,x)=({L_0^{[-\jj-1]}(\mu)L_0^{[\jj+1]}(x)})^{-1}=(\mu-i\frac{\jj+1}2)^{-1}(x+i\frac{\jj+1}2)^{-1}.
\eeq
is the normalization constant. We often omit arguments if $\mu=\xi$ e.g. $\LL_\jj\equiv\LL_\jj(\mu,\mu)$
All these operators act on $V_\jj\ox V_\jj$, where $V_\jj$ is the module of the representation $\jj$  spanned by $e_k,\, (k=0,...\jj)$. Useful facts are:

\ben
\- Charges are invariant under: (a) cyclic shift of nodes, (b) interchange $1\leftrightarrow 2$.
\- For each node: $[(\LL_\jj(\mu,\xx))^i_i,\hat S^z]=0,\; i=1,2 $ (no sum), where $\hat S^z=s_\jj^z\ox\one +\one\ox s_\jj^z$. We decompose $V_\jj\ox V_\jj$ as direct sum of eigenspaces of $S^z$:
$V_\jj\ox V_\jj=\oplus_{S^z=0}^{2\jj} W(\jj,S^z)$. Thus
 $\MM_\jj^\psi(\mu,x): W(\jj,S^z)\to W(\jj,S^z)$.
\-  $\vv_\jj\in  W(\jj,\jj)$
\een
For simple $\psi$ one can easily obtain the left unit eigenvector $w$ \refeq{eigenv}: 
\beqa
(e_k\ox e_{\jj-k})\cdot(\LL_\jj)^1_1&=&
\frac{\norm}{4} \left[((2 k + 1-\jj)^2 + 4 \mu^2 ) e_k\otimes e_{\jj-k}+4 k (k-\jj-1) e_{k-1}\otimes e_{\jj-k+1}\right]
\non
(e_k\ox e_{\jj-k})\cdot(\LL_\jj)^2_2&=&  \frac{\norm}{4} \left[ ((2 k - 1-\jj)^2 + 4 \mu^2 ) e_k\otimes
   e_{\jj-k}+4 (k+1) (k-\jj) e_{k+1}\otimes e_{\jj-k-1}\right]\nn
\eeqa
where $\norm(\jj)=((1+\jj)^2/4+\mu^2)^{-1}$. Then:
\beq
 \ww_\jj=\sum_{k=0}^{\jj} (-1)^k\,e_k\ox e_{\jj-k},\; i=1,2,
\eeq
It follows that $\ww_\jj\in W(\jj,\jj)$
and also $\vv_\jj\in W(\jj,\jj)$ what significantly simplifies calculations of charges.
\subsection{Poles of $X_1$}\label{app:X1}
Here we shall determine alignment of poles of $X_1$.

From $\ww^\dagger({\LL})^i_i=\ww^\dagger,\; i=1,2$ (no sum) the 2$\times$2 matrix $\MM$ has the form
\beq
\MM=\twomat ab{a-1}{b+1}
\eeq 
then 
\beq
\ww^\dagger \vv=N\, \frac{a+b-1}{1-a}=N\, \frac{\det(\MM)-1}{1-a},\quad 
\eeq
where $N$  is a normalization constant.
Vanishing of the numerator: $\det(\MM)-1=a+b-1=0$ is the condition for $x=1$ to be double zero of \beq\det(\MM-x)=0. 
\eeq
When additionally $a=1$ (i.e. also $b=0$) then $\MM$ has two eigenvalues equal 1. Thus
$\ww^\dagger \vv=0$ is the condition for the $\MM$ to have non-trivial Jordan form.
From  $\det(\LL_{1,ii}(\mu))=\frac{\mu^2}{\mu^2+1}$ one gets $\det(\MM)-1=(\frac{\mu^2}{\mu^2+1})^M-1=0$. Substituting $\mu^2=y-1/2$ 
we obtain\footnote{We have excluded $k=0$ because it corresponds to $y\to i\infty$ limit which is not seen for finite $M$.}
\beq\label{curve}
\frac{y-1/2}{y+1/2}=e^{2i\pi k/M},\quad  k=1,...M-1
\eeq
that means that $y\in i \RR$. One must remember, though, that not all the solutions of \refeq{curve} are poles of $X_1$, but certainly all these poles align
the hyperbola: $\Im(\mu)^2-\Re(\mu)^2=1/2$. This fact can be seen on  Fig.\ref{fig:X-ch-7} and Fig.\ref{fig:X-40-1}.



\subsection{Derivation of \refeq{approx}}
\label{app:large-mu}

We discuss derivation of  \refeq{approx} which well approximate 
 charges in PS. We do kind of hybrid $1/mu$ expansion in which the normalization factor $\norm$ is kept intact. 

We are looking for leading and the first subleading term of $\vv$ and $\MM$
(subscript \jj{} is mostly skipped here):
\beq
\MM=\prod_{i\in\psi} (\LL)^i_i
\eeq
 in $|\mu|\to\infty$ expansion. We shall expand terms from Lax operators only.
 The normalization factor $\norm$ will be left intact.
The following observations are helpful: 
\bit
\item the diagonal elements of $\MM$ contain the leading terms. These are $(\mu^\pm\pm is^z)\ox(\mu^\pm\pm is^z)$;
\item the off-diagonal terms $s^\pm\ox s^\mp$ are always suppressed;
\item  $s^\pm\ox s^\mp$ can be freely shifted along the chain because their commutator with
$(\mu^\pm\pm is^z)\ox(\mu^\pm\pm is^z)$ is ${\cal O}(\mu^{0})$ i.e. suppressed by two powers of $\mu$.
\eit
In this way we get
\beqa
\MM(e_k\ox e_{\jj-k})&\approx&[1-\frac1{\mu^2}(k^2 M+k(n_1-n_2-\jj M)-\jj n_1)]
\, e_k\ox e_{\jj-k}\\
&&-\frac{n_1}{\mu^2} (\jj-k)(k+1)\,e_{k+1}\ox e_{\jj-k-1}-\frac{ n_2}{\mu^2} (\jj-k+1)k\,(e_{k+1}\ox e_{\jj-k-1})\nn
\eeqa
where $n_1,\ n_2$ denotes numbers of spins up and down in $\psi$.
From the above one easily gets:
\beq\label{v0}
\vv\approx \sum_{k=0}^\jj (-1)^k r^k\, (e_k\ox e_{\jj-k}),\quad \ww^\dagger\vv\approx \frac{1-r^{\jj+1}}{1-r}
\eeq
where $r=n_2/n_1$. 
Notice that $\ww^\dagger\vv$ is spectral parameter $\mu$  independent contrary to exact results on charges. Solutions to $\ww^\dagger\vv=0$ give spectra of the bound states which we should not  expect to appear at $\mu\to \infty$ limit, at least in the leading order. Thus $\ww^\dagger\vv=const$ is physically well motivated.

In similar manner we calculate $\p\MM$. 
Derivatives $\p_\xx\LL$ are proportional to $(\mu^\pm\pm is^z)\ox 1$ 
which can be shifted to back of all expressions at the 
cost of commutators. The latter are higher order corrections, thus irrelevant here. Hence $\p\MM$ contains a sum of expressions of the form
\beq\label{m1-p}
\prod_{i\in\psi'}\LL^i_i\cdot  (\mu^\pm\pm is^z)\ox \one
\eeq
where $\psi'$ is a subchain in which one node  (where the derivative acted) was removed.
Because finally we are interested in $\ww^\dagger \p\MM\, \vv$, due to $\ww^\dagger\LL^i_i=\ww^\dagger$ the $\LL$'s in \refeq{m1-p} can be omitted yielding
\beq\label{m1-a1}
\p\MM\approx \norm\left[\,n_1((\mu^-+ is^z)\ox \one )
+n_2((\mu^-- is^z)\ox \one)\,\right]-\frac{M}{\mu+i(\jj+1)/2}
\eeq
where the last term comes from differentiation of the normalization $\norm$
: $\p_x \norm(\mu,x)|_{x=\mu}$. Now we can use $\vv$ displayed in
\refeq{v0} to get our final result \refeq{approx}.
\beq\label{app:approx}
X_\jj(\mu)\approx\frac1{2\pi}\frac1{\mu^2+\frac{1}{4} (\jj+1)^2}\left(\frac{\jj}{2}-\frac{\jj (1-r) \left(r^{\jj+1}+1\right)-2 r \left(1-r^\jj\right)}{2
   (r+1) \left(1-r^{\jj+1}\right)}\right)
\eeq

It is worth to notice that nontrivial denominator comes from $\norm$ of \refeq{LL}.
 The $\frac{1}{4} (\jj+1)^2$ piece regularizes behaviour of $X_\jj(\mu)$ for small $\mu$. 
%

\newpage
\subsection{More pictures}
\label{app:pic}
In this section we show several additional pictures which help to understand 
the main paper.
\vskip-.2cm
\begin{figure}[h]
\centering
\caption{Poles of $X_2(\mu)$ for $\psi=\{1, 1, 1, 1, 2, 1, 2\}$ displayed in the complex half-plane ($\Im(\mu)\geq 0$) and the corresponding analytic expression below.}
\includegraphics[width=5cm]{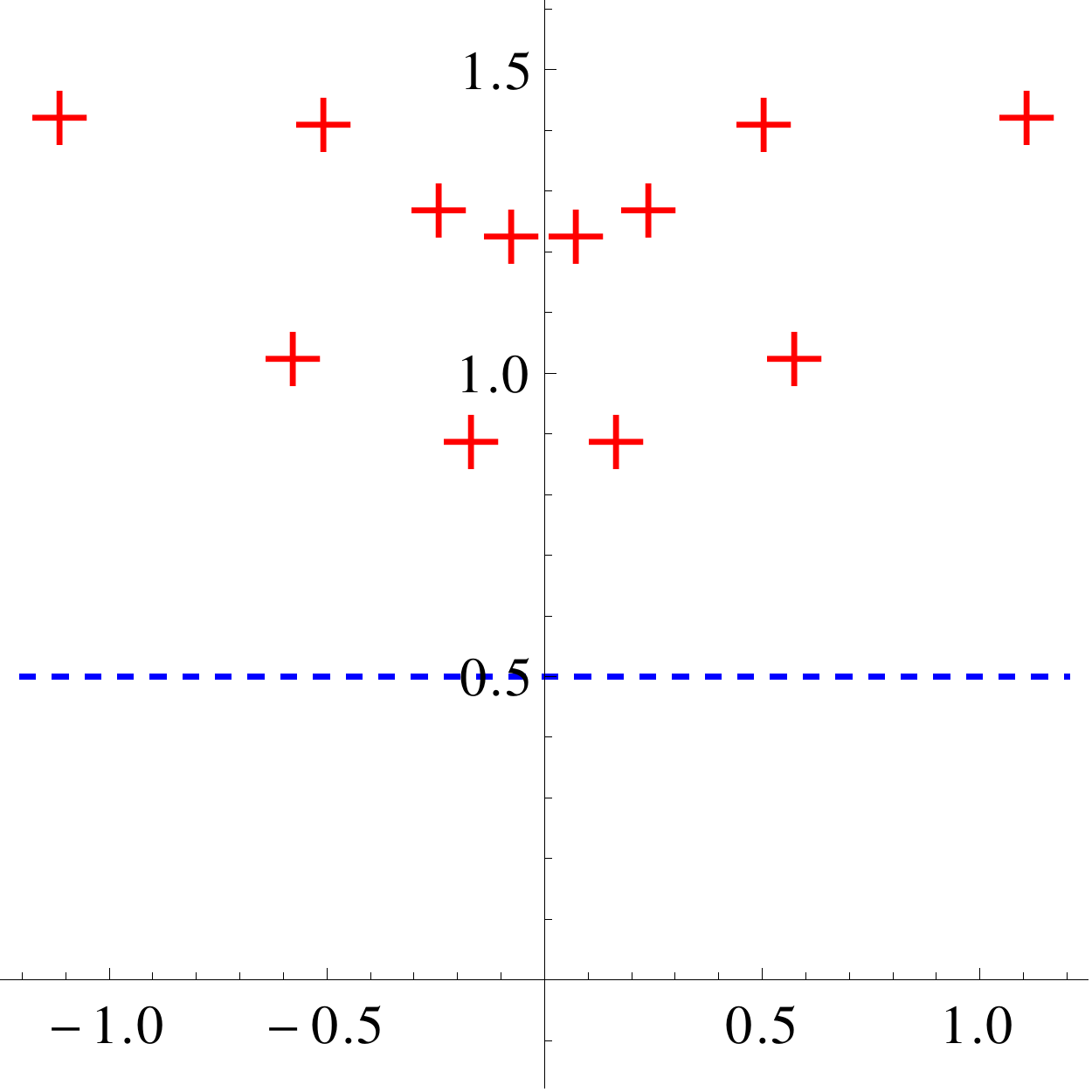}
\label{fig:X-ch7-2}
\vskip-0.8cm
{\scriptsize \beqa
X_2(\mu)=\frac{12}{7\pi}\times \hfill\hspace{15cm} &&\non
 \frac{146800640
   \mu^{22}+1871708160 \mu^{20}+12689080320 \mu^{18}+57839910912 \mu^{16}+189502291968
   \mu^{14}+455242522624 \mu^{12}}
   {654311424
      \mu^{24}+9479127040 \mu^{22}+71485620224 \mu^{20}+360711192576 \mu^{18}+1319572668416
      \mu^{16}+3603429982208 \mu^{14}+7414633218048 \mu^{12}}&&
  \non
  \frac{ +\quad 804831242240 \mu^{10}+1039513800192 \mu^8+958560474048
      \mu^6+599204434384 \mu^4+227327105092 \mu^2+39573895547}{+\; 11483489935360
   \mu^{10}+13232857409792 \mu^8+11037736083712 \mu^6+6304816157920 \mu^4+2204519902544
   \mu^2+356177462887}&&\nn
\eeqa}
\caption{Distribution of poles of $X_2$  and $X_3$ for the substate $\psi$ of the length $M=40$:
$\psi=\{1,2,1,1,1,2,2,1,1,1,1,2,1,2,1,1,2,1,1,1,1,1,2,1,2,1,2,1,2,2,1,2,2,2,1,2,1,1,2,1\}$ .} 
\vskip.2cm
\includegraphics[width=6cm,keepaspectratio]{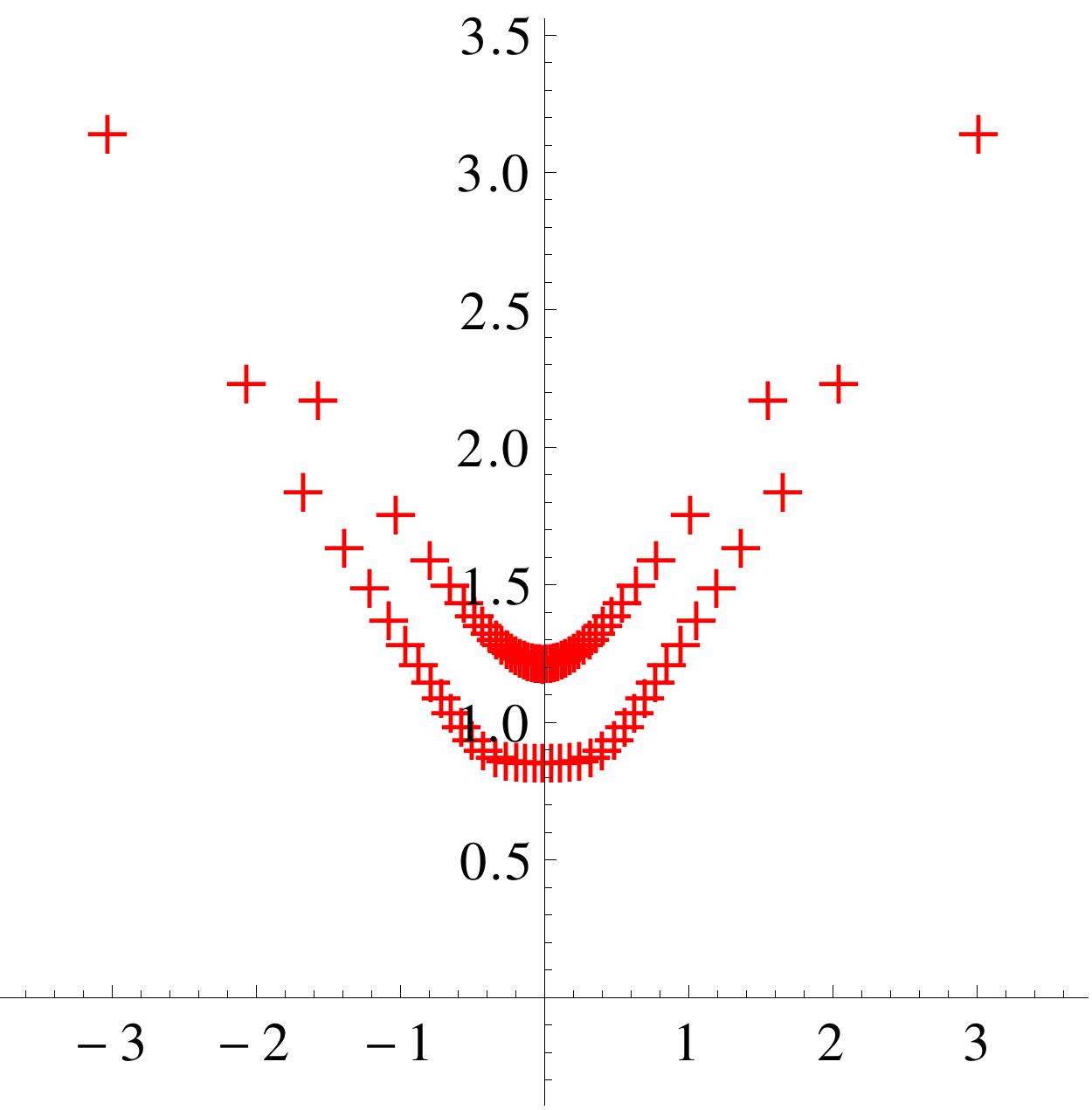}\hskip2cm
\includegraphics[width=6cm,keepaspectratio]{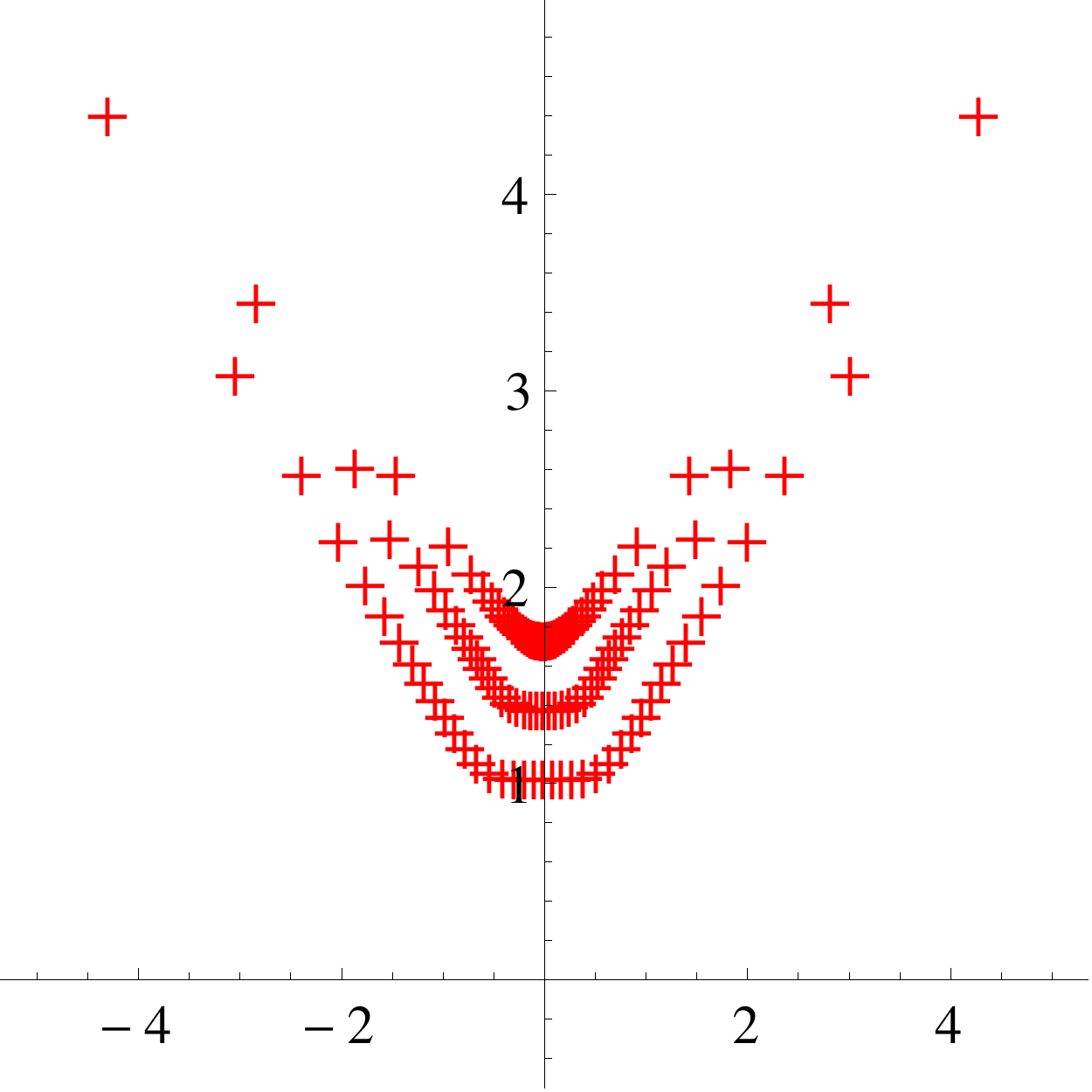}
	\label{fig:X-40-23}
\end{figure}

\newpage
\begin{figure}
\caption{Below we present several figures comparing $X_\jj(\mu)$ (red dashed lines) and $\tilde{X}_\jj(\mu)$ (black lines) of \refeq{approx} for real $\mu\in [-10,10]$. Separate figures on the right show the relative difference $\ndis_\jj$ given by
\refeq{ndis}. The displayed cases are $\jj=1,4,8$ for $\psi=\{1,1,1,2,1,2,2,1,2,1\}$.}\vskip1cm
\centering
\includegraphics[width=6.5cm]{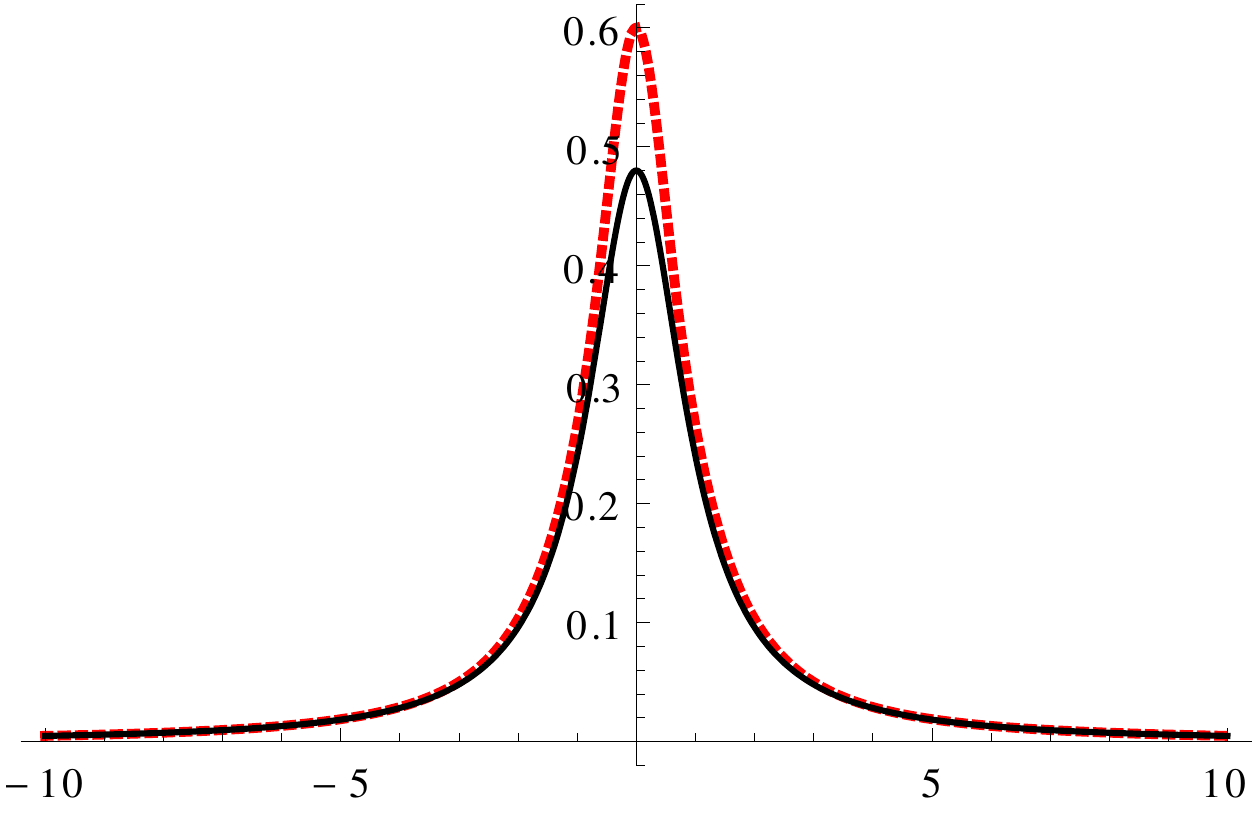}\hskip1cm
\includegraphics[width=6.5cm]{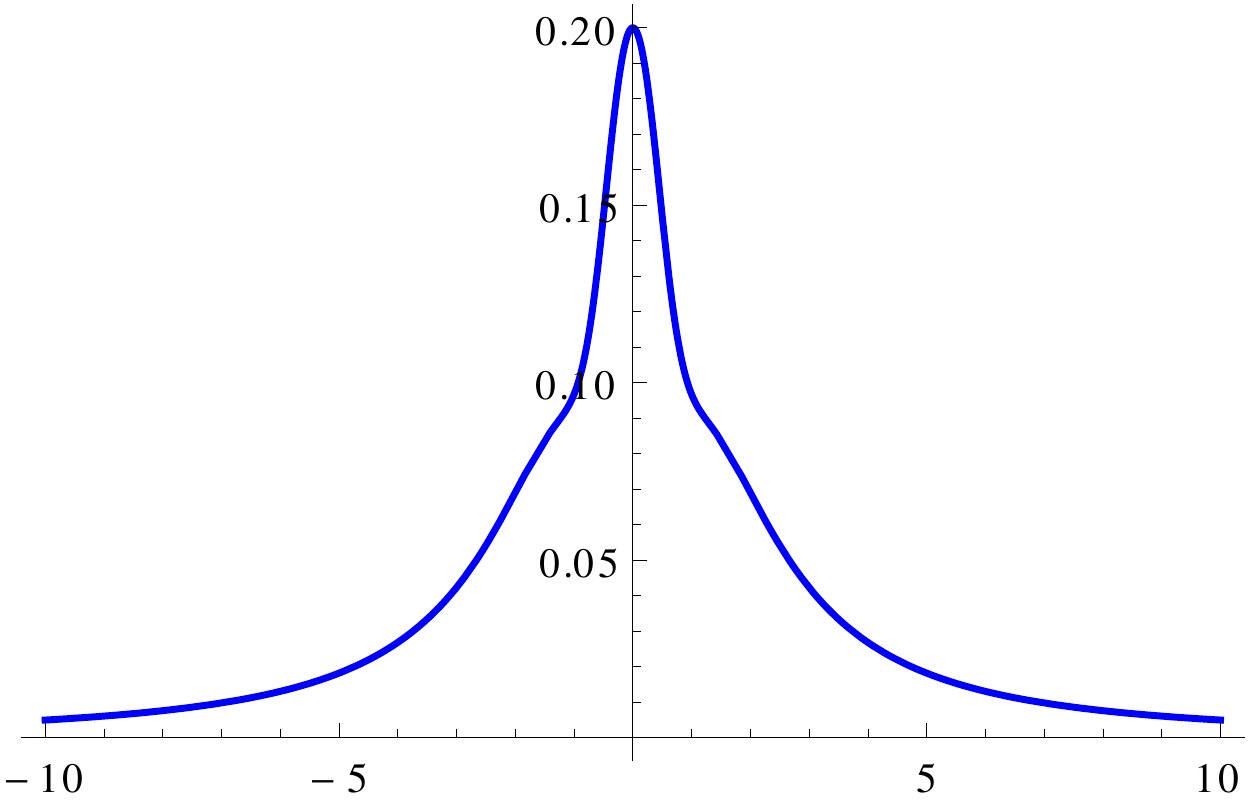}
\label{fig:ndis}
\vskip.5cm
\includegraphics[width=6.5cm]{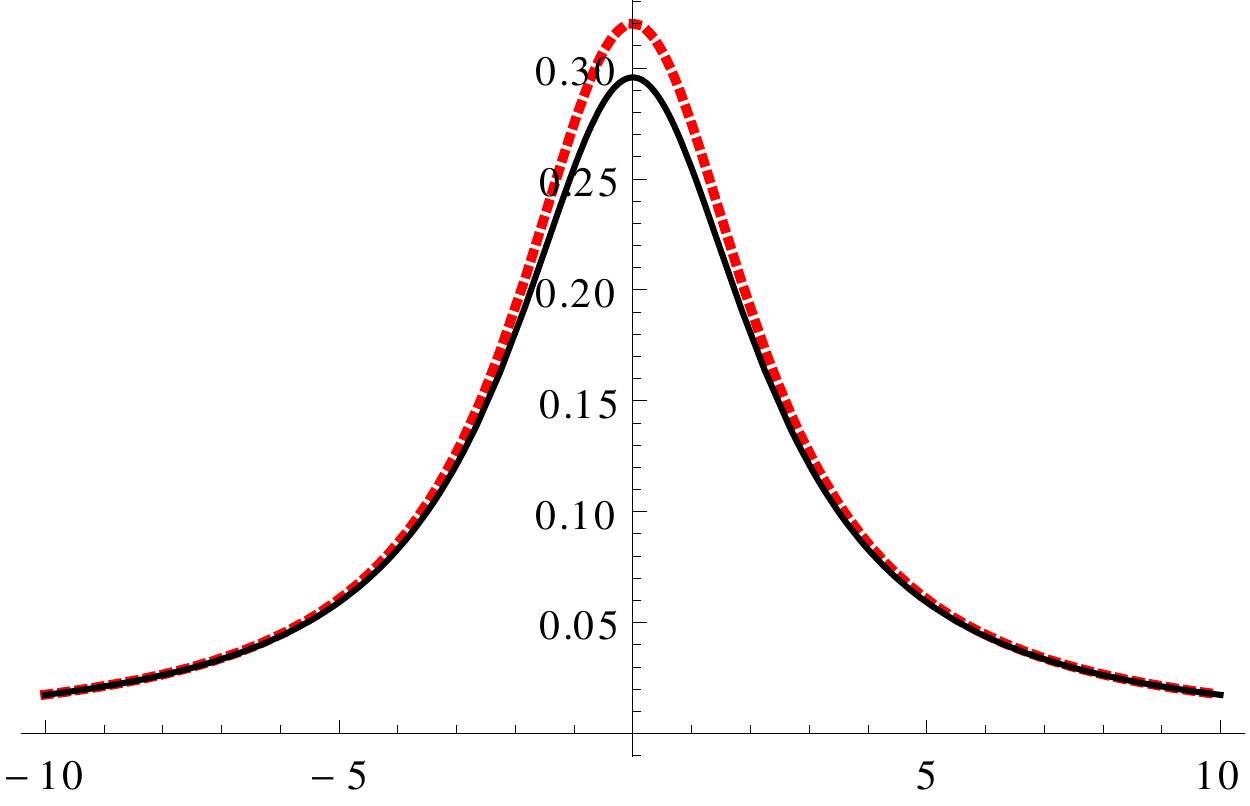}\hskip1cm
\includegraphics[width=6.5cm]{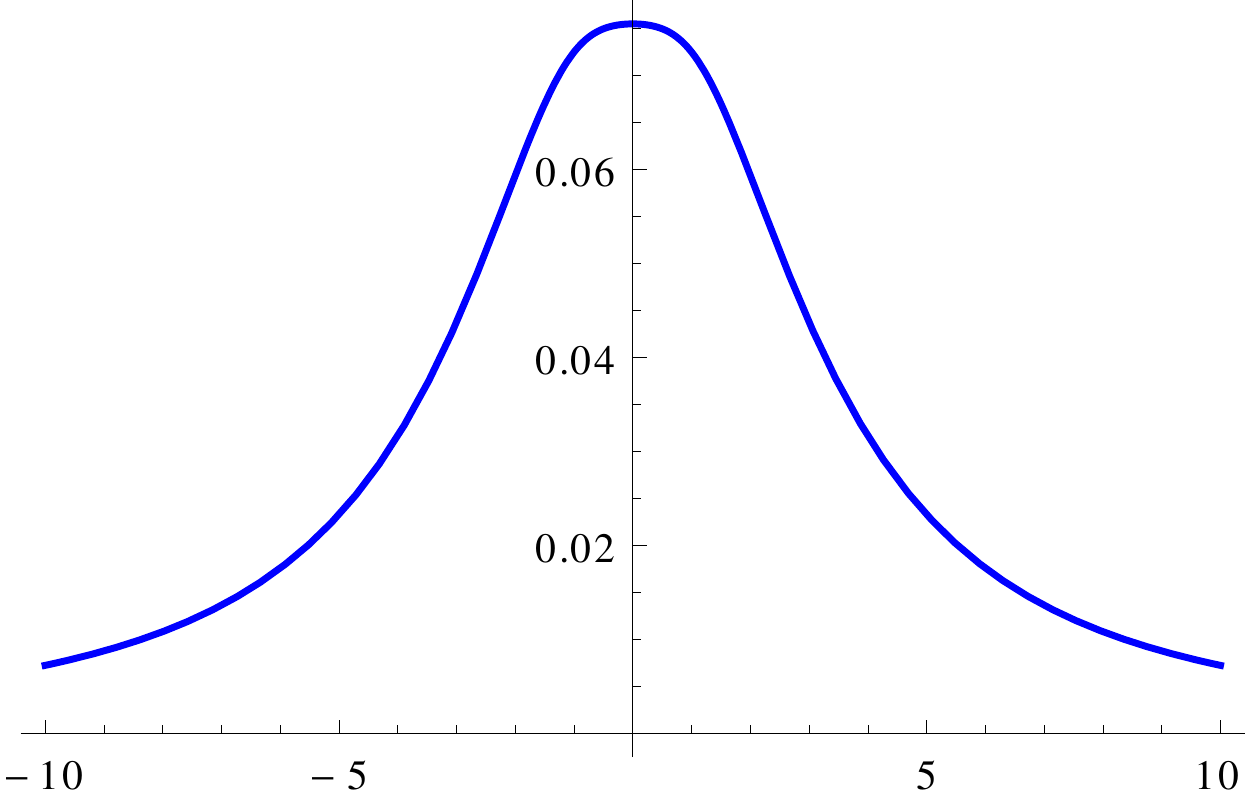}\\
\vskip.5cm
\includegraphics[width=6.5cm]{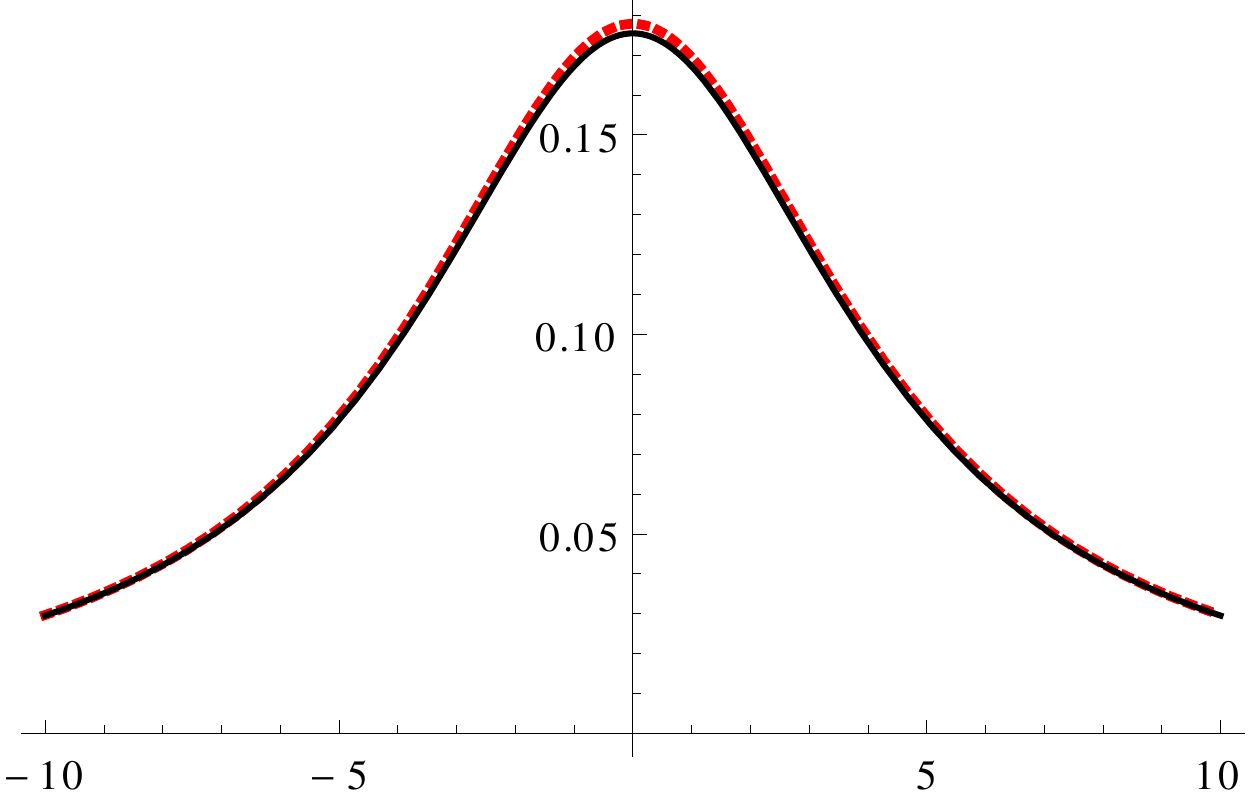}\hskip1cm
\includegraphics[width=6.5cm]{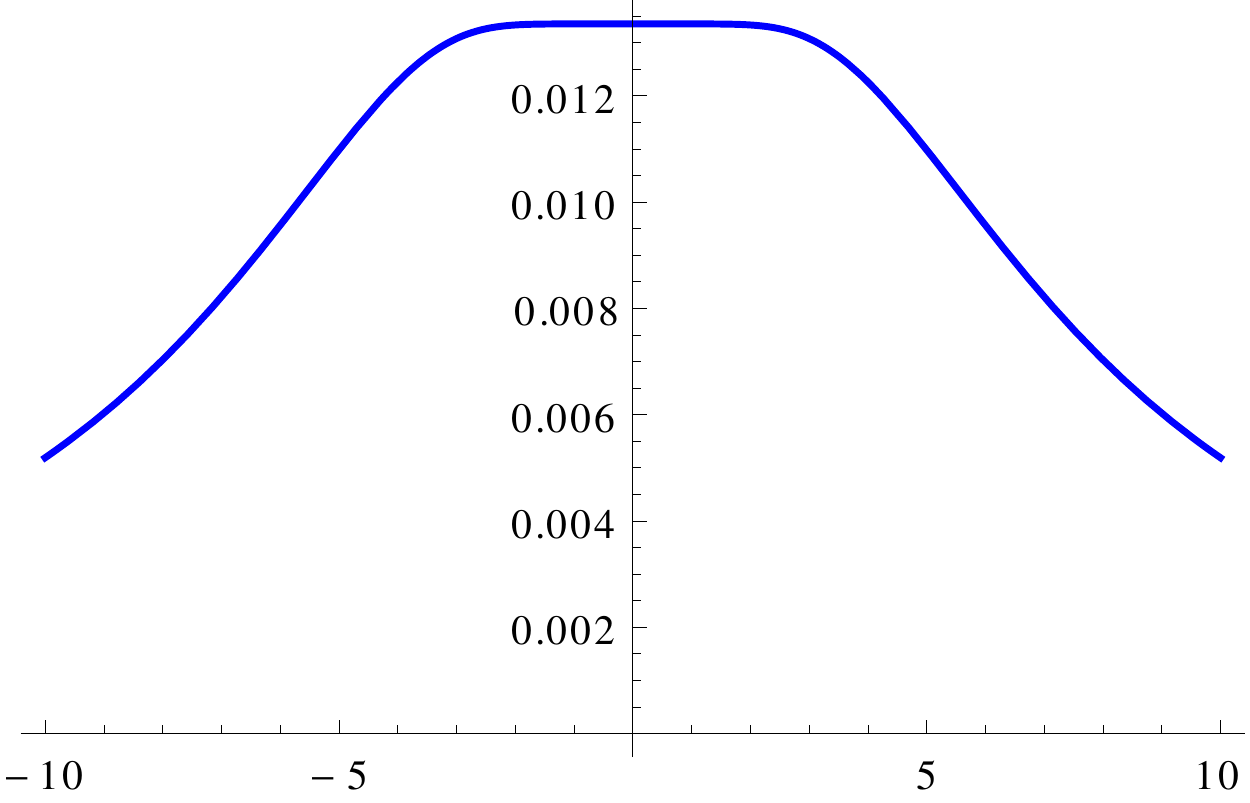}\\
\end{figure}

\end{document}